# Disappearing errors in a conversion model

by
David P Fan, University of Minnesota, USA
dfan@umn.edu

*Abstract*

The same basic differential equation model has been adapted for time-dependent conversions of members of a population among different states. The conversion model has been applied in different contexts such as epidemiological infections, the Bass model for the diffusion of innovations, and the ideodynamic model for public opinion. For example, the ideodynamic version of the model predicts changes in public opinions in response to persuasive messages extending back to an indefinite past. All messages are measured with error, and this chapter discusses how errors in message measurements disappear with time so that predicted opinion values gradually become unaffected by past measurement errors. Prediction uncertainty is discussed using formal statistics, sensitivity analysis and bootstrap variance calculations. This chapter presents ideodynamic predictions for opinion time series about the Toyota car manufacturer calculated from daily Twitter scores over two and half years. During this time, there was a sudden onslaught of bad news for Toyota, and the model could accurately predict the accompanying drop in favorable Toyota opinion and rise in unfavorable opinion.

*Introduction*

This chapter considers a longitudinal conversion model which is applicable to any population which can be divided into proportions with members in each proportion belonging to a separate state. The model is explored for epidemiological infections, the diffusion of innovations and public opinions.

Longitudinal models can be stochastic or mechanistic with hybrids also possible. In their discussion of epidemiology, (Holmdahl & Buckee, 2020) use the term mechanistic for mathematical functions which reflect the logical mechanism of a system. Different systems can have different structures and processes so different systems can lead to quite different mechanistic equations. Error behavior can be difficult to analyze for mechanistic functions with arbitrary structures.

Obstacles in error management have led to a stochastic approach which can be justified using Taylor series to expand any time-dependent mechanistic function to give a polynomial with an infinite number of terms. The expansion can be made around any point in time with the only requirement being that the mechanistic function should be infinitely differentiable. A useful feature of the Taylor series is that lower degree terms are more dominant at times closer to the expansion point.

Therefore, it is not uncommon for stochastic models to approximate some unknown mechanistic function with a Taylor series and then truncating the series to just the linear term assuming that interest is focused on behaviors of time-dependent variables close to the expansion point. Such a truncation facilitates error management because errors in linear terms have been studied extensively. However, the loss of higher degree terms means that the linear approximation can result in increasingly serious deviations from the underlying mechanistic function as time departs from the expansion time.

Rather than beginning with a truncated polynomial approximation, this chapter discusses a mechanistic conversion model which is customized to arguments about the system. The logic leads to a differential equation model which predicts proportions of the population in various states over time. The proportions are modeled to change in response to conversion forces which will be discussed.

Trajectories of population proportions in the conversion model are computed by integrating differential equations over time with conversion forces being the drivers of change. If the conversion forces are measured with error, then errors in the conversion forces will accumulate during the integration process. One possibility is that error accumulations would lead to predictions with ever greater uncertainties over time as happens in hurricane forecasts with their continually expanding cones of uncertainty.

Fortunately, the opposite happens. The error calculations in this chapter show how early errors in the predicted population proportions disappear to give predicted time series that are mainly affected by errors from the recent past. The formal statistics are bolstered by sensitivity analyses and bootstrap variance calculations.

This chapter will introduce the model using a minimal mechanism for epidemiological infections. Then, the model will be extended to the diffusion of innovations. From there, the model will be broadened further to include predictions of public opinions from mass media messages which are measured with error.

*Epidemiological infections*

A precedent for infection models is the logistic differential equation introduced by Verhulst in 1838 to describe the growth of a population (Bacaër, 2011). The differential equation and its variants, have since been used extensively to predict trajectories of infections during epidemics (Gao & Hethcote, 1992). This infection approach (Chowell et al., 2019) has appeared in a number of papers including those (Haushofer & Metcalf, 2020; Holmdahl & Buckee, 2020; Roosa et al., 2020; Vannabouathong et al., 2020) for the spread of COVID-19 infections.

For a minimal infection model, consider a system with a closed population with a fixed set of members. The members will be divided into two compartments or proportions with one proportion consisting of members in infected state $I$ due to infection by a microbe. The remaining proportion will have members who have not yet been infected and are therefore in susceptible state $S$. At any time $t$, population members in state $I$ will belong to proportion $y_I(t)$, and all members in state S will belong to proportion $y_S(t)$. All proportions of the total population must sum to one at all times so

$$y_I(t) + y_S(t) = 1 \tag{1}$$

The next step in the argument recognizes that it is difficult to measure the physical presence of the infectious microbes which constitute the actual forces for infection. Therefore, all infected individuals in proportion $y_I(t)$ are assumed to have the same chance of releasing infectious microbes to give an infectious conversion force $F_I(t)$ with

$$F_I(t) = k y_I(t). \tag{2}$$

Change in infection $dy_I(t)/dt$ at time $t$ requires both infectious force $F_I(t)$ and susceptible individuals in proportion $y_S(t)$. All individuals in $y_S(t)$ are assumed to have the same chance of being infected by $F_I(t)$ so the two terms are multiplied together to give

$$\frac{dy_I(t)}{dt} = F_I(t) y_S(t). \tag{3}$$

With the substitution of $y_S(t)$ from (1) and $F_I(t)$ from (2), (3) becomes the logistic differential equation

$$\frac{dy(t)}{dt} = k y(t)[1 - y(t)] \tag{4}$$

after dropping all subscripts $I$. This differential equation integrates explicitly to the logistic function $y(t) = 1/(1 + e^{-kt})$. This minimal infection equation is often modified for actual epidemiological applications by including other considerations such as infected members entering the recovered state and ceasing their release of microbes.

Despite its simplicity, major arguments about population conversions are expressed in this introductory model. The model describes time-dependent interconversions among population members in mutually exclusive states. For infections, the two states are susceptible and infected. The conversions in general are due to conversion force functions which act to move members from specific target subpopulations to specific destination subpopulations. In the minimal infection model, infectious forces can only transfer susceptible individuals to the infected state; infectious forces should not recruit any individuals from the infected proportion because these individuals have already been infected. There is one conversion equation for each population proportion, with one of the equations explicitly stating that all population proportions must sum to one. For infection, the two equations are (1) and (3).

In the logistic function, the only predictor for infection at time $t$ is $t$ so there is no need to consider errors arising from measurements at other times.

### *Diffusion of innovations*

The (Bass, 1969) conversion model for diffusion is an extension of the logistic infection model. The model was based on observations by Rogers (Rogers, 1962) that the increase in the adoption of an innovation can sometimes also have the S-shape of the logistic function. For diffusion, Bass proposed a persuasive conversion force function

$$F_G(t) = k_G[p + q y_G(t)] \tag{5}$$

with subscript $G$ referring to a good innovation to be adopted. In (5), the second term on the right, $qy_G(t)$, is a force of imitation that has the same form as the infectious force in (2). This force corresponds to all adopters of the innovation in proportion $y_G(t)$ broadcasting favorable information with a constant probability given by constant $q$. In addition to this force of imitation, Bass adds a presumptively constant force for innovation represented by constant $p$. The force of innovation can originate from any external source favoring the innovation including advertising. The total of the forces from both innovation and imitation are multiplied by persuasibility constant $k_G$ in (5) to give the final $F_G(t)$. Substitution of this force into the equivalent of (3) for infection gives the Bass model for the diffusion of innovations

$$\frac{dy_G(t)}{dt} = F_G(t)[1 - y_G(t)] = k_G[p + qy_G(t)][1 - y_G(t)]. \tag{6}$$

In the special case of no information of innovation, $p = 0$ so that (6) for innovations becomes (4) for infections with the only difference being the added subscript $G$ and the absorption of constant $q$ into the constant of proportionality. Thus, the elimination of constant $p$ leads to the logistic increase in adoption described by Rogers.

### Ideodynamic model

The Bass model can not only be simplified to give a logistic time series, but the same model can also be extended to yield the ideodynamic model for predicting time series of public opinion (D. P. Fan, 1985, 1988). The extension is accomplished using two modifications.

*Persuasive force.* The first modification for ideodynamics notes that the only variable predictor in Bass persuasive force $F_G(t)$ in (5) is the proportion of adopters $y_G(t)$. The ideodynamic model takes the alternate approach of computing $F_G(t)$ from empirically measured mass media messages scored to favor the good opinion. The assumption is that the mass media can represent all persuasive information including that coming from adopters. Mass media messages are indiscriminately broadcast and hence should be generally available throughout the population. With the assignment of the mass media messages as sufficient sources for all persuasive information, the ideodynamic persuasive conversion force is

$$F_G(t) = k_G M_G(t) \tag{7}$$

where $M_G(t)$ includes all good media messages used by the public at time $t$.

The computation of $M_G(t)$ at time $t$ begins with empirically measured $m_G(t)$, the total volume of good media messages arriving at the population at time $t$. Besides $m_G(t)$, $M_G(t)$ also includes memories of $m_G(t')$ from all prior times $t'$. Memory is assumed to decay exponentially over time with a constant decay rate $r$ so $M_G(t)$ includes additive contributions $m_G(t')e^{-r(t-t')}$ from all messages arriving at earlier times $t'$. The totality of the messages used by the population includes contributions beginning at time $t$ and extending back to the first available messages at time $t_0$ so

$$M_G(t) = k_{M,G} \int_t^{t_0} m_G(t')e^{-r(t-t')}dt' \tag{8}$$

where $k_{M,G}$ is the constant of proportionality accounting for the inherent persuasibility of favorable messages as well as for the use of a sample rather than all relevant persuasive media messages.

*Persuasive force errors.* In practice, values for $m_G(t')$ are estimated from a representative sample rather than all good mass media messages. Sample measurements $\widehat{m_G}(t')$ are likely to be reasonably constant over time so that $\widehat{m_G}(t')$ can be assigned a Poisson random error $\epsilon_G$ which is approximately the same at all times so

$$m_G(t') = s_G(\widehat{m_G}(t') + \epsilon_G) \tag{9}$$

with constant of proportionality $s_G$. Substitution of (9) into (8) and then (5) gives

$$F_G(t) = k_G \int_t^{t_0}(\widehat{m_G}(t') + \epsilon_G)e^{-r(t-t')}dt'. \tag{10}$$

where constant $s_G$ is absorbed into $k_G$.

Error term $\delta_G(t) = k_G \int_t^{t_0} \epsilon_G e^{-r(t-t')}dt'$ from the second term on the right of (10) integrates explicitly to $\delta_G(t) = \frac{k_G \epsilon_G}{r}(1 - e^{-r(t-t_0)})$. In this expression, error $\delta_G(t)$ increases along with time gap $t - t_0$ before reaching the stable value of

$$\delta_G \to \delta_G(\infty) = \frac{k_G \epsilon_G}{r} \tag{11}$$

when $t$ becomes very large relative to $t_0$. Consequently, early errors disappear leaving the error in $F_G(t)$ to depend only on recent errors in measured messages.

*Population conversions.* The second modification of the Bass model for the ideodynamic model recognizes that a good opinion can not only be adopted as a result of good persuasive force $F_G(t)$ but can also be unadopted in response to bad messages with subscript $B$ to give bad persuasive force $F_B(t)$ with the same structure as $F_G(t)$.

Inclusion of unadoption leads to the extension of (3) for infection and (6) for the Bass model to give the ideodynamic model

$$\frac{dy_G(t)}{dt} = F_G(t)[1 - y_G(t)] - F_B(t)y_G(t). \tag{12}$$

where the $F_G(t)$ is given by (8). The second term $-F_B(t)y_G(t)$ on the right is prefixed by a minus sign to indicate that $F_B(t)$ acts to decrease good opinion in proportion $y_G(t)$. Differential equation (12) integrates explicitly to

$$y_G(t) = \frac{\int F_G(t)e^{\int(F_G(t)+F_B(t))dt}dt + c}{e^{\int(F_G(t)+F_B(t))dt}} \tag{13}$$

where $c$ is the constant of integration. This integrated solution shows that the trajectory of good option $y_G(t)$ over time depends entirely on persuasive force functions $F_{(\cdot)}(t)$.

*Conversion errors.* Success with the use of (13) depends on predicted $y_G(t)$ having errors with stable variances despite errors in the measured $F_{(\cdot)}(t)$.

The following error calculation is a minor modification of the presentation in Appendix B of (D. P. Fan & Cook, 2003). The computation begins by defining $u_{G,0} \equiv F_G/(F_G + F_B)$ and $v \equiv exp(\int (F_G + F_G)dt)$ so that integrating (13) by parts gives

$$y_G(t) = \frac{\int u_{G,0}\,dv + c}{v} = \frac{u_{G,0}v}{v} - \frac{\int u_{G,1}\,dv + c}{v} \quad (14)$$

where $u_1 \equiv (du_{G,0}/dt)/(F_G + F_G)$. The integration by parts can continue iteratively to give

$$y_G(t) = u_{G,0} - u_{G,1} + u_{G,2} \ldots u_{G,n} \ldots + c/v \quad (15)$$

where $u_{G,n} \equiv (du_{G,n-1}/dt)/(F_G + F_B)$ and the $u$ terms are preceded by alternating plus and minus signs.

Note that $v$ can only increase with integration time because $v$ is the integral of all prior conversion forces which are derived from message scores that are all inherently positive numbers. Consider the situation where enough time has elapsed so that $v$ is sufficiently large compared to $c$ that $c/v$ can be ignored. Under this condition, (15) has no remaining $v$ terms. With no integral terms, prediction error $\Delta_G(t) \equiv \hat{y}_G(t) - y_G(t)$ only reflects current values of $u$ and its derivatives and is independent of the integration. Therefore, predicted time series (13) has the useful feature that $var(\Delta_G(t))$ only depends on local values of $u_{G,n}$. This feature makes it practical to compute $\hat{y}_G(t)$ from $\hat{F}_{(\cdot)}(t)$ functions alone without the worry that $var(\Delta_G(t))$ will increase without limit as $t \to \infty$.

If $\hat{y}_G(t)$ changes only slowly and within a narrow range, then $u_{G,0}$ is the dominant term in (15) because all other terms $u_{G,n}$ are based on derivatives which are small when $u$ changes little. Note that the condition of unchanging $u_{G,0}$ only requires that the ratio $F_G/(F_G + F_B)$ should not vary without the more stringent condition of constancy in the individual $F_{(\cdot)}$ functions. Given that the $F_{(\cdot)}$ are associated with additive, unbiased errors $\delta_{(\cdot)}$ from (11), then the first approximation is $E(\Delta_G) = 0$ and

$$\begin{aligned}
var(\Delta_G) &\approx var(u_{G,0}) \\
&= var(\delta_G)\left(\frac{\partial u_{G,0}}{\partial F_G}\right)^2 + var(\delta_B)\left(\frac{\partial u_{G,0}}{\partial F_B}\right)^2 \\
&= \frac{var(\delta_G)F_B^2 + var(\delta_B)F_G^2}{(F_G + F_B)^4}
\end{aligned}$$

if enough time has elapsed that the effects of the constant of integration $c$ have dissipated due to increasing $v$. This equation assumes that $F_G$ is independent of $F_B$. Since (15) leads to stability in $var(\Delta_G)$, $\hat{y}_G(t)$ can be predicted using any numerical method.

*Sensitivity analysis.* Aside from formal statistics, sensitivity analysis is another means for obtaining insights into error behavior. In sensitivity analysis, errors are deliberately introduced into calculations to see how they affect predictions.

In implementing (13), errors are usually minimized by initializing the integration with opinion values that are close to the initial time of calculation $t_0$. Sensitivity analyses can be performed by purposely using erroneous initial opinion values that are far from any actual opinion measurements. One set of initial values with maximum errors would be $y_G(t_0) = 1$ and $y_B(t_0) = 0$; another would be the reverse with $y_G(t_0) = 0$ and $y_B(t_0) = 1$.

The error calculation given above indicates that early error should disappear so predictions beginning with extremely different initial conditions should converge to the same value, and this result has actually been found (D. P. Fan & Cook, 2003).

*Bootstrap variance.* Yet a third way to obtain insights into errors is to compute bootstrap variances. A bootstrap calculation starts with a presumptively representative sample drawn from a full population. This starting sample is then used as a substitute full population that is presumed to have a structure close to that of the original full population. Since inferences can be drawn from representative samples, the bootstrap method draws random samples of the substitute full population. The random draws are made with replacement to arrive at redrawn samples of the same size as the substitute population. Statistics are then calculated from the redrawn samples (Hall, 1994).

In a bootstrap calculation for (13), 200 random draws with replacement were made of the full set of message scores for the entire length of the time series. Each draw had the same number of scores as in the original set of scores. Then, predictions were made using the model parameters estimated from the original dataset. Variances computed from the replicate predictions again showed that the variances in the predicted opinions were stable over time (D. P. Fan & Cook, 2003).

## I*deodynamic application to data

For ideodynamic regressions, it is more convenient to perform the integration iteratively using differential equation (12) than to calculate integral equation (13). For the iterative computation, (12) can be converted to discrete time to give the nonlinear regression equation:

$$y_{G,t} = y_{G,t-1} + k_G M_{G,t}(1 - y_{G,t-1}) - k_B M_B(t) y_{G,t-1} \qquad (16)$$

Implementation of (16) begins at initial time $t = 0$ and an initial $y_{G,0}$ which can be an opinion measurement taken close to the beginning of the computation. Then iterative computations are made by incrementing time one step after another to predict the entire time course from persuasive information alone.

It is straightforward to extend the ideodynamic methodology to predict simultaneous conversions among more than two opinion states. This chapter illustrates an extension of the model to predict the time courses of three opinions about Toyota Motor Corporation from 6/15/2009 to 8/1/2012 (D. Fan et al., 2013).

The predictions were for opinions from the BrandIndex program of YouGov plc, London, UK. Responses were collected daily using an internet survey question asking, "Do you have a generally positive feeling about the brand?" Responses were averaged weekly to give opinion values with sample sizes of around 650. Respondents were assigned to time $t$ in the middle of the week to give the three opinion proportions of positive $y_{Pos,t}$, negative $y_{Neg,t}$, and neutral $y_{Neut,t}$ for Toyota.

The persuasive force predictors came from Twitter sentences scored daily by General Sentiment, Inc. of Long Island, New York using the Lydia computer system (Key et al., 2010). Daily scores for good and bad Toyota sentences were summed to give positive $M_{Pos,t}$ and negative $M_{Neg,t}$ persuasive messages.

Prior findings from (D. P. Fan & Cook, 2003) and elsewhere have shown consistently that the exponential decay of the memory of persuasive messages is so rapid that media messages effectively cease to change opinions after a single day. Longer memories of old information would result in more sluggish responses of opinions to new information than is found in regressions studies. Instead, opinion responds so nimbly to incoming information that old information must be forgotten very rapidly.

Therefore, the Toyota model assumed that each sentence was only persuasive on its day of appearance and had vanished by the following day.

The volume of total Tweets increased quite dramatically during the two and a half years of the study so all positive and negative Twitter scores in a day were normalized by dividing by the total volume of scores on that day.

The study was conducted from 2009 to 2012 because Toyota had a strong reputation until 8/28/2009 when a California Highway patrol car crashed and killed the driver, an off-duty officer, and his family. This event led to Toyota's 9/29/2009 recall of 3.9 million vehicles for uncontrollable acceleration due to floor mats that caused accelerator pedals to stick. Besides a further recall of 2.3 million cars due to accelerator issues, Toyota had other problems leading to further recalls including those of its bestselling Prius Hybrid for braking problems. The company recalled a total of 8 million cars in 2009 and 2010. Media content about Toyota subsided after these spikes of bad publicity. Toyota provided an excellent case for testing the model under circumstances where persuasive information was not stationary during the time course but underwent a large and unanticipated perturbation instead.

The survey data divided the public into the three proportions of $y_{Pos,t}$, $y_{Neg,t}$, and $y_{Neut,t}$ so the ideodynamic model was configured to account for transfers of opinion-holders among the three groups.

Among potential conversion patterns, those in Fig. 1 gave satisfactory predictions. In the model, people would first be persuaded to move from positive to neutral with enrollment in the negative group occurring in a subsequent step requiring more negative information. At the same time, reverse transitions could also occur with people moving from the negative to the neutral group and from there to holding a positive opinion.

Since people convert from their previous opinions $y_{(\cdot),t-1}$ at time $t-1$ one day earlier to their present opinions $y_{(\cdot),t}$ at time $t$, every force function $F_{(\cdot),t} = k_{(\cdot)} M_{(\cdot),t}$ from (7) is multiplied by a target $y_{(\cdot),t-1}$ and its associated persuasibility constant $k_{(\cdot)}$. This multiplication makes the model nonlinear. The result is (17) and (18) below. In addition, (19) derives from the requirement that all proportions must sum to one.

$$y_{Pos,t} = y_{Pos,t-1} + k_{Pos2} \times M_{Pos,t} \times y_{Neut,t-1} - k_{Neg1} \times M_{Neg,t} \times y_{Pos,t-1} \quad (17)$$
$$y_{Neg,t} = y_{Neg,t-1} + k_{Neg2} \times M_{Neg,t} \times y_{Neut,t-1} - k_{Pos1} \times M_{Pos,t} \times y_{Neg,t-1} \quad (18)$$
$$y_{Neut,t} = 1 - y_{Pos,t} - y_{Neg,t} \quad (19)$$

Model (17-19) is implemented in the accompanying documentation and Excel files which provide all the opinion data from BrandIndex's internet surveys and Twitter scores from General Sentiment. The iterative opinion prediction was initialized with Twitter data on 6/15/2009 and measured opinion from 6/17/2009, just two days later. The closeness in these times means the initial opinions were likely to be reasonably accurate representations of opinion at the start of the iterative opinion calculation. The Excel files also provide regressions and statistics for (17-19) performed using the Excel add-in module called Solver (see online appendix).

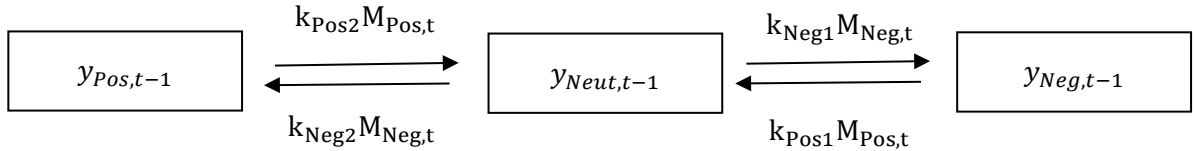

Fig. 1: Model for conversions among proportions $y_{(\cdot)}$ of a population holding opinions positive, neutral and negative for the Toyota Corporation. The conversions were from time $(t-1)$ to time $t$ in response to persuasive information $F_{(\cdot),t} = k_{(\cdot)} M_{(\cdot),t}$ at time $t$.

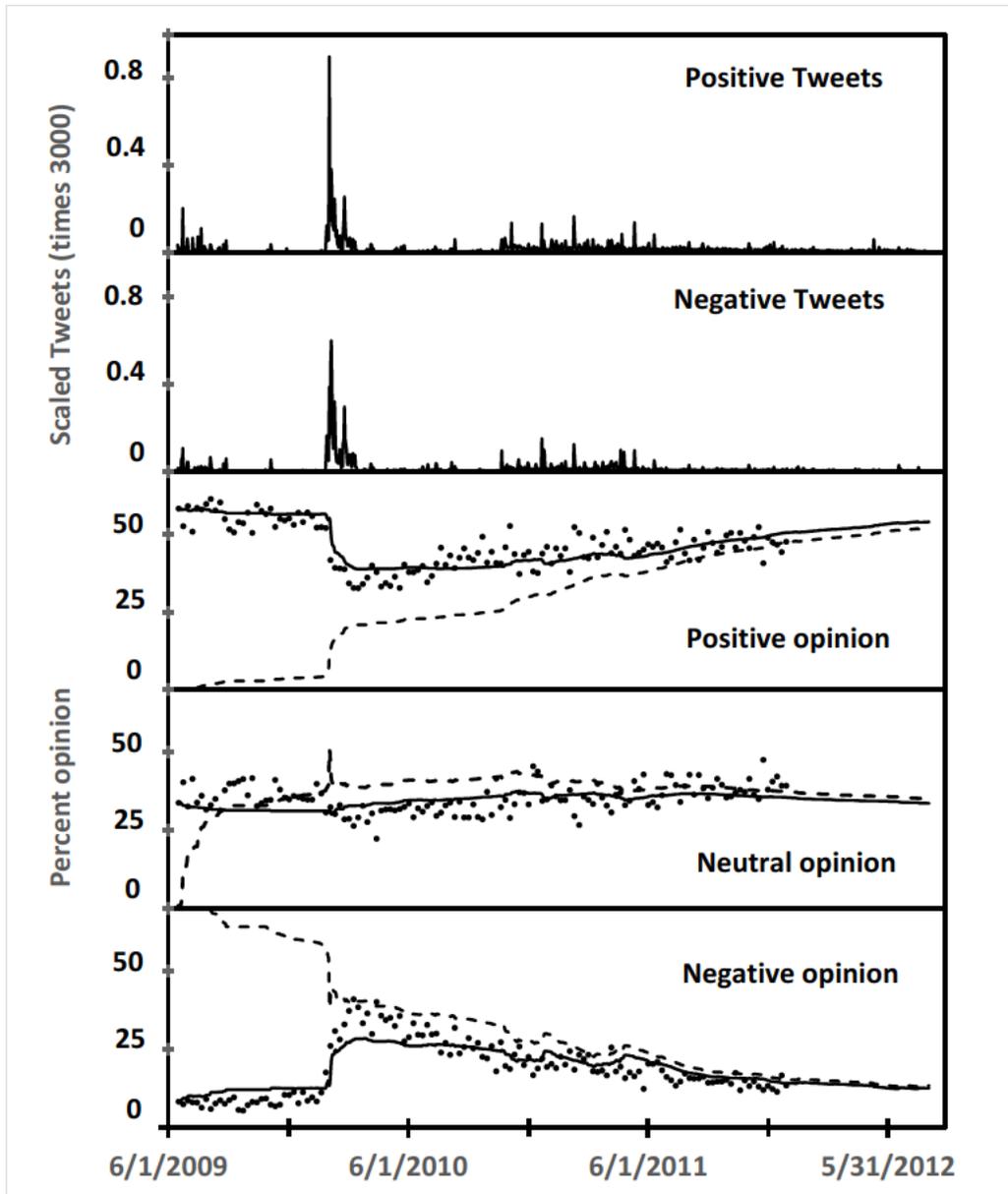

Fig. 2. Toyota Tweets and opinion. The top two frames give daily positive (top frame) and negative (second frame from top) Tweets scored by General Sentiment scaled as the proportion of all Tweets on a given day; all proportions are multiplied by 3000 for ease of display. The third to fifth frames from the top give positive (third frame), neutral (fourth frame), and negative (fifth frame) opinions about Toyota. The dots show opinions averaged by the week from BrandIndex. The solid lines show ideodynamic predictions initialized with measured opinions taken two days after the beginning of the iterative computations. The dashed lines show the same ideodynamic predictions initialized with the extreme opinion values of $y_{Neg,0} = 1$ and $y_{Neut,0} = y_{Pos,0} = 0$.

The spikes in both positive and negative Twitter content (Fig. 2, top two frames) corresponded to the timing of the Toyota accelerator revelations and their aftermaths.

These Tweet spikes led to the steep drop in positive opinions (Fig. 2, third frame, solid line) and the accompanying rise in negative perceptions (Fig. 2, bottom frame, solid line). Neutral opinion was on the transition pathways from positive to negative opinion and back (Fig. 2, fourth frame, solid line) and moved very little.

The model only required the estimation of four coefficients which maintained great stability as seen in the closeness of the predicted line to measured opinion values throughout the time course using parameters estimated from opinion measurements extending to the end of 2011 (Fig. 2). This stability allowed for plausible opinion predictions from Twitter data for the next eight months.

Tweets are assumed to represent relevant persuasive information used by the public rather than the entirety of that information. Therefore, the absolute values of the estimated persuasibility constants for Tweets are not meaningful. Instead, more useful numbers are the relative values of the constants. Therefore, the estimated coefficients were all normalized to $k_{Pos2}$ to give the persuasive powers of their associated persuasive force functions. The normalized coefficients were all approximately the same with $k_{Pos2} = 1$, $k_{Neg1} = 1.1$, $k_{Neg2} = 2.5$, and $k_{Pos1} = 3.1$. The similarity of the coefficients meant that both positive and negative Tweet scores could all represent the positive and negative persuasive information about equally well.

The quality of the fit is seen in the root mean squared error of 4.3 percent which is in the range of survey error. Also, the prediction gave the high R-squared values of 0.69 for $y_{Pos,t}$ and 0.84 for $y_{Neg,t}$. The low R-squared value of 0.02 for $y_{Neut,t}$ is expected because the course line for $y_{Neut,t}$ is practically flat, so no model should give much improvement over a horizontal line.

In the calculations discussed so far, the initial values for the iterative predictions were only two days later than the initial time of the calculations. Therefore, these values should have had minimal errors. For a sensitivity analysis, the effect of early errors was explored by beginning the iterative calculations with the extreme erroneous initial conditions of $y_{Neg,0} = 1$ and $y_{Neut,0} = y_{Pos,0} = 0$. The parameters estimated from the low error initial conditions were used for the iterative calculations of the sensitivity analysis. As expected from the formal error calculations, opinion predictions beginning with the highly erroneous initial values (Fig. 2, bottom three frames, dashed lines) did converge to predictions using the more accurate initial conditions (Fig. 2, bottom three frames, solid lines).

Thus, the sensitivity analysis confirmed the statistical calculations showing that predicted opinion trajectories should gradually come to depend only on recent messages and their errors.

*Discussion*

The mechanistic conversion model in this chapter is applicable to epidemiological infections, the Bass model for the diffusion of innovations and time series of public opinions. The flexibility of the model derives from the division of the model into two separate steps. The first step is to calculate applicable conversion forces; the second is to compute the response of a population to the forces.

Consider first the conversion forces. For infections, the actual conversion forces should be microbes themselves. However, empirical measurements of infectious particles are often impractical, so the forces are assumed to be proportional to the proportion of infected individuals. Using similar arguments, persuasive forces in the Bass diffusion model also include a component of imitation that is proportional to the subpopulation of adopters. This component is augmented by a mathematical constant to account for a presumptively constant external force of innovation.

Studies more recent than the 1969 Bass model have used empirically measured media messages in the ideodynamic model in place of the mathematical assumptions of the Bass model. The replacement was made because changes in behaviors should depend on persuasion by information as happens for public opinion. The replacement was also enabled by the more recent availability of mass media content scored by computer from electronic databases. Examples of satisfactory predictions from the mass media include those for the adoption of such behaviors as the purchase of caffeine-free colas and the usages of cocaine, tobacco, or the National AIDS Hotline (D. P. Fan, 1993, 1996b, 2002; D. P. Fan & Elketroussi, 1989; D. P. Fan & Holway, 1994; D. P. Fan & Shaffer, 1990).

For these and other ideodynamic predictions based on mass media information, the model proposes an exponential decay for the memory of the information in the persuasive forces. The error calculations for the conversion forces show that the exponential decay leads to opinions changing largely in response to information remembered from the very recent past.

Error calculations for persuasive forces showed the disappearance of early errors in the force functions given exponential memory losses. The calculations were performed in continuous time to take advantage of the fact that integrals of exponentials are the same exponentials.

Moving on to the conversion process, mathematically assumed conversion forces can give S-shaped logistic increases for both infections and Bass diffusion. The mathematically modeled forces predict logistic trajectories which are not straight lines in time. In contrast, linear models would give approximations that are only appropriate over short time periods around individual points in a time series.

Unlike the infection and Bass models where entire predicted time series depend on constant parameters, the ideodynamic model uses measured conversion forces as the sole determinants of any conversions. If there are no conversion forces, ideodynamics predicts that opinion would stay unchanged.

Empirically measured persuasive forces can both increase and decrease the proportion of people holding an opinion. Over time, the proportion of individuals with an opinion becomes the result of all past opinion changes. Thus, opinion at a particular time can be interpreted as a memory which summarizes all of a person's past opinion adoptions even if the person had forgotten when or how the changes were made.

As was the case for errors in force functions, error calculations for the conversion equations were also performed in continuous time so that integration by parts for integral equations could give easy computations to show error disappearance. Standard methods for stochastic differential equations (Oksendal, 2013) were not needed for the error calculations for either the force functions or the conversion functions.

The disappearance of early errors from statistical calculations was supported by sensitivity analyses in which large errors were introduced into the initial conditions of the iterative opinion predictions. Results from the Toyota study and from (D. P. Fan & Cook, 2003) both indicated that errors in prior opinions vanish over time.

The third procedure for examining errors was to compute bootstrap variances, and these calculations also demonstrated the error stability found from formal statistics and sensitivity analyses.

There are two keys to the disappearance of early errors in the conversion equations. One key is the logical requirement that population proportions sum to one at all times. Another key comes from the argument that persuasive forces only act on relevant target proportions. This requirement is implemented by multiplying conversion forces by target proportions throughout the iterative calculations. The multiplication leads to continuous decreases in contributions by persuasive forces over time because population proportions always lie between zero and one.

The exclusion of prior opinion measurements from opinion predictors is the opposite of the approach in autoregressive models like ARIMA where essential predictors are earlier values of the same opinion time series. Furthermore, autoregressive models tend to include only variables coming from a limited number of past time intervals. In contrast, the ideodynamic model includes all information back to the beginning of the iterative computations. The error stability allowed the same parameters and news media coverage to make successful predictions of the University's Index of Consumer Sentiment every day over two and a half decades (D. P. Fan & Cook, 2003).

Furthermore, the ideodynamic model provides critical tests of both the appropriateness of the model and the suitability of the input message scores which are the only sources for the predictors. Successful ideodynamic predictions of Toyota opinions could be made from not only Twitter scores (see above) but also content scores from the mainstream press, blogs, and internet forums (D. Fan et al., 2013). Therefore, all these media have similarly predictive content, and the model is also appropriate.

The independence of the media predictors from the opinions being predicted allows the two types of variables to be measured at different times and to have missing data. Twitter scores were generally scored every day. However, there were a few days when data were unavailable. When there are no additional scores, the ideodynamic model simply adds no new information to the persuasive forces. If this omission leads to opinion prediction errors, then the errors gradually disappear.

The decoupling of force and opinion measurements also meant that opinions could be predicted daily from Twitter scores and then be compared in regressions to opinion measurements whenever the opinions were measured, however infrequently or irregularly. The sample sizes for Toyota opinion measurements were only about 100 per day, so it was necessary to average opinions by the week to obtain sample sizes that were large enough to avoid excessive scatter in the opinion values. Therefore, daily predicted opinion values were compared with opinion measurements once a week.

The ability to use ideodynamics in regressions with irregularly timed opinion data was found in a variety of applications which demonstrated that the mainstream press could give consistently satisfactory predictions of opinion time courses in the United States (Domke et al., 1997; D. P. Fan, 1996a; D. P. Fan et al., 2002; D. P. Fan, Wyatt, et al., 2001; D. P. Fan & Norem, 1992; D. P. Fan & Tims, 1989; Hertog & Fan, 1995; Huebner et al., 1997; Jasperson et al., 1998; Jasperson & Fan, 2002; Shah et al., 1999). Ideodynamics could also predict opinion time series in a variety of other countries including Canada (D. P. Fan, 1997), Germany (D. P. Fan, Brosius, et al., 2001), and the Netherlands (Kleinnijenhuis & Fan, 1999)

Ideodynamic models might even be applicable in an era of microtargeted information. In microtargeting, reinforcing information is often directed preferentially to people already favoring an opinion. Such reinforcing information should not alter opinions; instead, opinion change requires information favoring an idea not held. In effect, opinion changes should mainly result from non-microtargeted information.

Although the ideodynamic model has been consistently successful for predicting public opinions and behaviors, the model has not been widely adopted. One contributory explanation could have been the rarity in the past of time series of computer-generated media scores for use as predictor variables as well as time series of opinions and behaviors for use as dependent variables. However, both these types of data have become much more readily available and can be supplemented from even newer data sources such as SocialScienceOne. This is a collaborative project for data transfer from major commercial companies like Facebook to the academic community.

<div style="text-align: center;">*References*</div>

Bacaër, N. (2011). Verhulst and the logistic equation (1838). In N. Bacaër, *A Short History of Mathematical Population Dynamics* (pp. 35–39). Springer London. https://doi.org/10.1007/978-0-85729-115-8_6

Bass, F. M. (1969). A New Product Growth for Model Consumer Durables. *Management Science*, *15*(5), 215–227. JSTOR.

Chowell, G., Tariq, A., & Hyman, J. M. (2019). A novel sub-epidemic modeling framework for short-term forecasting epidemic waves. *BMC Medicine*, *17*(1), 164. https://doi.org/10.1186/s12916-019-1406-6

Domke, D., Fan, D. P., Fibison, M., Shah, D. V., Smith, S. S., & Watts, M. D. (1997). News media, candidates and issues, and public opinion in the 1996 presidential campaign. *Journalism & Mass Communication Quarterly*, *74*(4), 718–737.

Fan, D., Geddes, D., & Flory, F. (2013). The Toyota Recall Crisis: Media Impact on Toyota's Corporate Brand Reputation. *Corporate Reputation Review*, *16*(2), 99–117. https://doi.org/10.1057/crr.2013.6

Fan, D. P. (1985). Ideodynamics: The kinetics of the evolution of ideas. *Journal of Mathematical Sociology*, *11*(1), 1–23.

Fan, D. P. (1988). *Predictions of public opinion from the mass media: Computer content analysis and mathematical modeling* (Vol. 12). Greenwood Publishing Group.

Fan, D. P. (1993). Quantitative estimates for the effects of AIDS public education on HIV infections. *International Journal of Bio-Medical Computing*, *33*(3–4), 157–177. https://doi.org/10.1016/0020-7101(93)90033-3

Fan, D. P. (1996a). Predictions of the Bush-Clinton-Perot presidential race from the press. *Political Analysis*, *6*, 67–105.

Fan, D. P. (1996b). Impact of the mass media on calls to the CDC National AIDS Hotline. *International Journal of Bio-Medical Computing*, *41*(3), 207–216. https://doi.org/10.1016/0020-7101(96)01196-8

Fan, D. P. (1997). Computer content analysis of press coverage and prediction of public opinion for the 1995 sovereignty referendum in Quebec. *Social Science Computer Review*, *15*(4), 351–366.

Fan, D. P. (2002). Impact of persuasive information on secular trends in health-related behaviors. *Public Health Communication: Evidence for Behavior Change*, 251.

Fan, D. P., Brosius, H.-B., & Esser, F. (2001). Computer and human coding of German text on attacks on foreigners. *PROGRESS IN COMMUNICATION SCIENCES*, 145–164.

Fan, D. P., & Cook, R. D. (2003). A differential equation model for predicting public opinions and behaviors from persuasive information: Application to the Index of Consumer Sentiment. *Journal of Mathematical Sociology*, *27*(1), 29–51.

Fan, D. P., & Elketroussi, M. (1989). Mathematical model for addiction: Application to Multiple Risk Factor Intervention Trial data for smoking. *Journal of Consulting and Clinical Psychology*, *57*(3), 456.

Fan, D. P., & Holway, W. B. (1994). Media coverage of cocaine and its impact on usage patterns. *International Journal of Public Opinion Research*, *6*(2), 139–162.

Fan, D. P., Keltner, K. A., & Wyatt, R. O. (2002). A matter of guilt or innocence: How news reports affect support for the death penalty in the United States. *International Journal of Public Opinion Research*, *14*(4), 439–452.

Fan, D. P., & Norem, L. (1992). The media and the fate of the Medicare Catastrophic Extension Act. *Journal of Health Politics, Policy and Law*, *17*(1), 39–70.

Fan, D. P., & Shaffer, C. L. (1990). Effects of the Mass Media News on Trends in the Consumption of Caffeine-Free Colas. *ACR North American Advances*, *NA-17*. http://acrwebsite.org/volumes/7016/volumes/v17/NA-17

Fan, D. P., & Tims, A. R. (1989). The impact of the news media on public opinion: American presidential election 1987–1988. *International Journal of Public Opinion Research*, *1*(2), 151–163.

Fan, D. P., Wyatt, R. O., & Keltner, K. (2001). The suicidal messenger: How press reporting affects public confidence in the press, the military, and organized religion. *Communication Research*, *28*(6), 826–852.

Gao, LindaQ., & Hethcote, HerbertW. (1992). Disease transmission models with density-dependent demographics. *Journal of Mathematical Biology*, *30*(7). https://doi.org/10.1007/BF00173265

Hall, P. (1994). Methodology and Theory for the Bootstrap. *Handbook of Econometrics*, *4*, 2341–2381.

Haushofer, J., & Metcalf, C. J. E. (2020). Which interventions work best in a pandemic? *Science*, *368*(6495), 1063–1065. https://doi.org/10.1126/science.abb6144

Hertog, J. K., & Fan, D. P. (1995). The impact of press coverage on social beliefs: The case of HIV transmission. *Communication Research*, *22*(5), 545–574.

Holmdahl, I., & Buckee, C. (2020). Wrong but Useful—What Covid-19 Epidemiologic Models Can and Cannot Tell Us. *New England Journal of Medicine*, NEJMp2016822. https://doi.org/10.1056/NEJMp2016822

Huebner, J., Fan, D. P., & Finnegan, Jr., John. (1997). "Death of a Thousand Cuts": The Impact of Media Coverage on Public Opinion About Clinton's Health Security Act. *Journal of Health Communication*, *2*(4), 253–270. https://doi.org/10.1080/108107397127581

Jasperson, A. E., & Fan, D. P. (2002). An aggregate examination of the backlash effect in political advertising: The case of the 1996 US Senate race in Minnesota. *Journal of Advertising*, *31*(1), 1–12.

Jasperson, A. E., Shah, D. V., Watts, M., Faber, R. J., & Fan, D. P. (1998). Framing and the public agenda: Media effects on the importance of the federal budget deficit. *Political Communication*, *15*(2), 205–224.

Key, E., Huddy, L., Lebo, M. J., & Skiena, S. (2010). Large scale online text analysis using Lydia. *APSA 2010 Annual Meeting Paper*.

Kleinnijenhuis, F., & Fan, D. P. (1999). Media coverage and the flow of voters in multiparty systems: The 1994 national elections in Holland and Germany. *International Journal of Public Opinion Research*, *11*(3), 233–256.

Oksendal, B. (2013). *Stochastic differential equations: An introduction with applications.* Springer Science & Business Media.

Rogers, E. M. (1962). *Diffusion of innovations.* Free Press of Glencoe, Free Press.

Roosa, K., Lee, Y., Luo, R., Kirpich, A., Rothenberg, R., Hyman, J. M., Yan, P., & Chowell, G. (2020). Real-time forecasts of the COVID-19 epidemic in China from February 5th to February 24th, 2020. *Infectious Disease Modelling*, *5*, 256–263.

Shah, D. V., Watts, M. D., Domke, D., Fan, D. P., & Fibison, M. (1999). News coverage, economic cues, and the public's presidential preferences, 1984-1996. *The Journal of Politics*, *61*(4), 914–943.

Vannabouathong, C., Devji, T., Ekhtiari, S., Chang, Y., Phillips, S. A., Zhu, M., Chagla, Z., Main, C., & Bhandari, M. (2020). Novel Coronavirus COVID-19: Current Evidence and Evolving Strategies.  [Miscellaneous Article]. *Journal of Bone*, *102*(9), 734–744. https://doi.org/10.2106/JBJS.20.00396